\documentstyle[12pt]{article}  
\begin{document} 

\title{Finding the Radiation Amplitude Zero in $W \gamma$ 
Production- Is it Unique to the Standard Model?}
\author{Mark A. Samuel \\ Dept. of Physics, Oklahoma State University, Stillwater, OK  74078 \\
Stanford Linear Accelerator Center, Stanford University, Stanford CA  94309 \\[2em]
and \\[2em] Tesfaye Abraha \\ 
Dept. of Physics, Oklahoma State University,  
Stillwater OK  74078}

\maketitle

\begin{abstract}
In the light of recent experimental observation of the Radiation 
Amplitude Zero (RAZ) in $W \gamma$ production by CDF at Fermilab, we 
consider its consequences. Is the RAZ unique to the Standard Model (S.M.)? 
Although it is not for $\bar{\nu}_e e^- \rightarrow W^- \gamma$, 
in the case of $d \bar{u} \rightarrow W^- \gamma$, which is the case 
of experimental interest, observation of the RAZ implies that the S.M. must 
be correct.
\end{abstract}

It is now 18 years since Radiation Amplitude Zeros were discovered by 
Mikaelian, Samuel and Sahdev, (1). These zeros which were found in the process
\begin{eqnarray}
d\bar{u} \rightarrow W^- \gamma     \nonumber  \\
\bar{\nu}e \rightarrow W^- \gamma   \nonumber  \\
u \bar{d} \rightarrow W^+ \gamma    \nonumber  \\ 	    
\nu e^+ \rightarrow W^+ \gamma	
\end{eqnarray}
were proposed as a means of measuring the magnetic moment of the $W$ 
boson, even though the $W$ was discovered only 4 years later.

These zeros are quite remarkable- the lowest-order amplitude vanishes 
for each spin state and the position of the zero is independent of the 
photon energy. (For massless quarks, it depends only on the quark 
charges) The RAZ provides a test of the magnetic moments of both the $W$ 
and the quarks; further, the position of the zero enables a direct 
measure of the fractional quark charges by real photons. It was later 
pointed out that the zero also occurs in radiative $W$ decay (2), where, in 
this case, the energy distribution vanishes along a certain line in the 
Dalitz plot. For related earlier work, see references (3) and (4).

Subsequently, (5-7) it was shown that these amplitude zeros can arise more 
generally, originating as the destructive interference of radiation 
patterns in gauge-theory tree amplitudes for massless gauge-boson 
emission. This is therefore a property of gauge theories; 
anomalous electromagnetic moments, for example, would spoil the perfect 
cancellations and such anomalies are forbidden in gauge couplings. For a 
specific analysis of the effect of $W$ anomalous moments in the $u\bar{d} 
\rightarrow W^+ \gamma$ reaction see ref (8). Of course, anomalous 
moments come up in higher-order corrections, and indeed RAZ do not appear 
beyond the tree approximation in any theory. For a careful assessment of 
QCD corrections see ref (9). For a generalization to more photons and
gluons, see ref (10)

Can we observe these zeros experimentally? A necessary condition is that 
all of the charges in both the initial and final states must be of the 
same sign or neutral. The best bet for observing RAZ experimentally is in 
the originally suggested sub-processes
\begin{eqnarray}
d\bar{u} & \rightarrow & W^- \gamma   \\  \nonumber
u\bar{d} & \rightarrow & W^+ \gamma
\end{eqnarray}
which can be seen in the processes
\begin{equation}
p\bar{p} \rightarrow W^\pm \gamma X
\end{equation}

A recent rapidity study by Baur {\em{et al}} (11) has given us a new and 
effective tool in the radiation zero analysis. They have shown that 
laboratory rapidity correlation involving the photon and the charged 
decay lepton display a pronounced dip corresponding to the RAZ.

Since it now seems certain that CDF at Fermilab will soon observe RAZ 
experimentally for the first time, a troubling thought occurs. {\em Is there an 
ambiguity? Could RAZ occur if $\kappa + \lambda =1$ even if we do not 
have a S.M. $W$ boson, $\kappa=1$, $\lambda=0$?} We will show that although 
such an ambiguity does exist for 
\begin{equation}
\bar{\nu_e} e^- \rightarrow W^- \gamma
\end{equation}
it does not occur for the experimentally relevant process
\begin{equation}
d\bar{u} \rightarrow W^- \gamma
\end{equation}
In the following we will consider only the 2 processes
$$\bar{\nu_e} e^- \rightarrow W^- \gamma$$
and
\begin{equation}
d\bar{u} \rightarrow W^- \gamma
\end{equation}
The corresponding results for $\nu_e e^+ \rightarrow W^+ \gamma$ and 
$u\bar{d} \rightarrow W^+ \gamma$ can easily be obtained from them.

We begin with the angular distribution
\begin{equation}
\frac{d \sigma}{dA}(q_i \bar{q_j} \rightarrow W^-\gamma)
= F(s,t,u,Q_i,\kappa, \lambda)
\end{equation}
where
\begin{eqnarray}
s+t+u &=& M_W^2 	\nonumber \\
t&=&-\frac{1}{2}(s-M_W^2)(1-\cos \theta)	\nonumber \\
u&=&-\frac{1}{2}(s-M_W^2)(1+\cos \theta)
\end{eqnarray}
and $\theta$ is the angle between the $W^-$ and the quark/electron in the 
center-of-mass frame. $Q_i$ is the charge of the quark/electron, and 
$\kappa$ and $\lambda$ are the parameters which determine the $W$ 
magnetic moment
\begin{equation}
\mu_W = \frac{e}{2M_W}(1+\kappa+\lambda)
\end{equation}
and its electric quadrupole moment
\begin{equation}
Q_W=-\frac{e}{M_W^2}(\kappa - \lambda )
\end{equation}
For the S.M., $\kappa=1$ and $\lambda=0$ and so
\begin{equation}
\mu_W = \frac{e}{M_W}
\end{equation}
\begin{equation}
Q_W = -\frac{e}{M_W^2}
\end{equation}

F is determined by crossing from 
\begin{equation}
\frac{d \sigma}{dA}(\gamma q_i \rightarrow q_j W^-)
\end{equation}
which is given in ref(12).

First we observe that for the S.M. ($\kappa=1$ and $\lambda=0$), the RAZ does indeed occur at
\begin{equation}
\cos \theta = -(1+2Q_i)
\end{equation}
We now set $\lambda=1-\kappa$, leaving $\kappa$ as a free parameter. We 
first  consider 
$\bar{\nu_e} e^- \rightarrow W^-\gamma$ where $Q_i=-1$ and  
$\cos \theta =1$. Here the RAZ is in the forward 
direction. By considering various values of s and $\cos \theta =1$ we find that
\begin{equation}
F(s,t(\cos \theta=1), u(\cos \theta=1),-1,\kappa, 1-\kappa)\equiv 0
\end{equation}
for any value of $\kappa$! Thus in this case we have an ambiguity. 
Observing  the RAZ in this case does not imply 
the S.M.! This is indeed disturbing.

We now turn to the experimentally important process
\begin{equation}
d\bar{u} \rightarrow W^-\gamma
\end{equation}
Here the RAZ occurs at $\cos \theta=-\frac{1}{3}$ as $Q_i=-\frac{1}{3}$. 
We  set $\cos \theta=-\frac{1}{3}$ and 
$\lambda=1-\kappa$. This time we find that
\begin{equation}
F(s,t(\cos\theta=-1/3),u(\cos\theta=-1/3),-1/3, \kappa, 1-\kappa) 
\propto  (1-\kappa)^2
\end{equation}
Thus in this case $F\neq 0$ unless $\kappa=1$ and, hence $\lambda=0$. 
Thus  nature is kind to us as the observation 
of a RAZ in this process requires the S.M. values $\kappa=1$ and 
$\lambda=0$  and, hence, means that the S.M. has 
survived another test beautifully!

\vspace{1em}
{\Large Acknowledgements}

It is a pleasure for M.A.S. to thank the theory group at SLAC for their 
kind  hospitality. He would also like to 
thank Stan Brodsky, Bob Brown, John Ellis, Steve Errede, Marek Karliner, Karnig 
Mikaelian and,  especially, Tom Rizzo, for valuable discussions. 

This work was supported by the U.S. Department of Energy under Grant 
Nos.  DE-FG05-84ER40215 and Contract No DE-AC03-76F00515.

\vspace{1em}
{\Large References}

\begin{description}
\item [1] K.O. Mikaelian, M.A. Samuel, and D. Sahdev, Phys. Rev. Lett. 43, 746 (1979) 
\item [2] T.R. Grose and K.O. Mikaelian, Phys. Rev. D23, 123 (1981) 
\item [3] K.O. Mikaelian, Phys. Rev. D17, 750 (1978) 
\item [4] R.W. Brown, D. Sahdev, and K.O. Mikaelian, Phys. Rev. D20, 1164 (1979) 
\item [5] S.J. Brodsky and R.W. Brown, Phys. Rev. Lett. 49, 966 (1982)
\item [6] M.A. Samuel, Phys. Rev. D27, 2724 (1983)
\item [7] R.W. Brown, K.L. Kowalski, and S.J.Brodsky, Phys. Rev. D28, 624 (1983)
\item [8] M.A. Samuel, N. Sinha, R. Sinha, and M.K. Sundaresan, Phys. Rev. D44, 2064 (1991)
\item [9] J. Smith, W.L. van Neerven, and J.A.M. Vermaseren, Z. Phys. C30, 621 (1986)
\item [] see also J. Smith, D. Thomas, and W.L. van Neerven, {\em ibid} 44, 267 (1989)
\item [] and S. Mendoza, J. Smith, and W.L. van Neerven, Phys. Rev. D47, 3913 (1993)
\item [10] R.W. Brown, M.E. Convery, and M.A. Samuel, Phys. Rev. D49, 2290 (1994)
\item [] see also the recent paper by U. Baur, T. Han, N. Kauer, R. Sobey, and D. Zeppenfeld, 
 $W\gamma\gamma$ Production at the Fermilab Teratron Collider: \\
  Gauge Invariance and Radiation Amplitude Zero, MADPH 97-986 (1997)
\item [11] U. Baur, S. Errede, and G. Landsberg, \\ 
 Proceedings of the Workshop on Physics at Current Accelerators and the SuperCollider, 
  Argonne, IL. (1993, unpublished)
\item [] see also U. Baur and D. Zeppenfeld, Nucl. Phys. B308, 127 (1988)
\item [] and U. Baur and E.L. Berger, Phys. Rev. D41, 1476 (1990)
\item [12] C.S. Kim, J. Lee, and H.S. Song, \underline{Investigation of $WW\gamma$ Couplings}, Invited talk at 
 Korea-Japan Joint Symposium, Dec. 1994. To be published by World Scientific, Singapore, edited by I.T. Cheon.
\end{description}

\end{document}